\documentclass[english]{article}
\usepackage[T1]{fontenc}
\usepackage[latin9]{inputenc}
\usepackage{geometry}
\usepackage{url}
\usepackage{hyperref}
\usepackage{amsmath}
\usepackage{amssymb}
\usepackage{graphicx}

\usepackage{cite}

\newcommand\rmd{\mathrm{d}}

\usepackage{babel}
\title{Resemblance of the power--law scaling behavior of a non--Markovian and nonlinear point processes}
\author{Aleksejus Kononovicius\thanks{email: \protect\href{mailto:aleksejus.kononovicius@tfai.vu.lt}{aleksejus.kononovicius@tfai.vu.lt};
website: \protect\url{http://kononovicius.lt}}, Rytis Kazakevi\v{c}ius, Bronislovas Kaulakys}
\date{Institute of Theoretical Physics and Astronomy, Vilnius University}
\begin{document}

\maketitle

\begin{abstract}
We analyze the statistical properties of a temporal point process
driven by a confined fractional Brownian motion. The event count distribution
and power spectral density of this non--Markovian point process exhibit
power--law scaling. We show that a nonlinear Markovian point process
can reproduce the same scaling behavior. This result indicates a possible
link between nonlinearity and apparent non--Markovian behavior.
\end{abstract}

\section{Introduction}

If someone knows the state of every particle in the universe, that
someone could determine the future of the universe from this single
glimpse at its current state. Therefore, if everything is known, the
universe would be Markovian (or even deterministic) to them. Though
our human awareness is often limited, the universe will often appear
non--Markovian to us \cite{VanKampen1998BJP}. The long--range memory
phenomenon seems to imply non--Markovian dynamics, yet we have successfully
used nonlinear Markovian models to reproduce it \cite{Kazakevicius2021Entropy}.
In this paper, we explore a link between nonlinearity and apparent
non--Markovian behavior by introducing and analyzing two point processes:
one is driven by a non--Markovian fractional noise and the other
by a Markovian white noise.

Many observed temporal phenomena in physical, biological, and social
systems are either discrete by their nature or continuous yet representable
as a discrete temporal pattern of instantaneous events \cite{Daley2008Springer}.
This temporal pattern can be encoded by a list of event times or inter--event
times. Numerous examples of phenomena described by event times or
inter--event times include contraction of infectious diseases \cite{Chen2021ResPhys},
order execution within financial markets \cite{Hambly2020AMF}, musical
rhythm \cite{Levitin2012PNAS}, earthquakes \cite{Pasari2017NatHaz},
traffic accidents \cite{Li2018RESS}, human heartbeat \cite{Kobayashi1982BioMed}
or formation of queues \cite{Bose2002Springer}. Such phenomena are
driven by various complex mechanisms with a degree of randomness and
causality (e.g., earthquakes trigger aftershocks). Such processes
can be modeled using the point process formalism \cite{Daley2008Springer}.

Poisson and Hawkes processes are two well--known examples of the
point processes. The Poisson point process assumes either a constant
event rate (homogeneous Poisson process) or time dependent event rate
(inhomogeneous Poisson process). While in the Hawkes process the system
effectively excites itself as newly occurring events temporarily increase
the event rate. Unlike in the Poisson process, in the Hawkes process
time increments are no longer independent. Having such embedded memory
property Hawkes process has found numerous applications in finance
\cite{Hawkes2017QF} and other fields \cite{Rizoiu2017FMR,Reinhart2018StatSci,Li2018RESS,Chen2021ResPhys}.
Recently it was shown that a certain class of nonlinear Hawkes processes
can reproduce power--law distributions \cite{Kanazawa2021PRL}, which
may appeal to practitioners dealing with complex systems exhibiting
power--law distributions, such as financial markets.

Long--range memory and $1/f^{\beta}$ noise (with $0.5\lesssim\beta\lesssim1.5$),
as its characteristic feature \cite{Mandelbrot1968SIAMR,Bak1987,West1989IJMPB,Ward2007Scholarpedia,Rodriguez2014PRE,Yadav2021},
are of particular interest as they are observed across different physical
\cite{Press1978CA,Dutta1981RMP,Balandin2013NN}, biological \cite{Kobayashi1982BioMed}
and social systems \cite{Cont2001RQUF}. Typically long--range memory
is modeled using various non--Markovian models: fractional Brownian
motion \cite{Decreusefond1999PA,Dieker2002Thesis,Benth2012,Beran2017Routledge,Akinlar2020CSF},
ARCH models \cite{Engle1982Econometrica,Bollerslev1986Econometrics,Engle1986EcoRev,Giraitis2000AAP,Giraitis2007,Giraitis2009}
or non--Markovian mechanisms such as trapping \cite{Chepizko2013PRL,Metzler2014PCCP}.
Our approach is different: in earlier works, we have used nonlinear
Markovian processes, such as point processes \cite{Kaulakys1998PRE,Kaulakys1999PLA,Kaulakys2005PhysRevE},
stochastic differential equations \cite{Kaulakys2004PhysRevE,Kaulakys2006PhysA,Gontis2010PhysA}
and agent--based models \cite{Kononovicius2012PhysA,Gontis2014PlosOne},
to model power--law statistical properties and long--range memory
phenomenon (for a recent review see \cite{Kazakevicius2021Entropy}).
In this paper, we explore how the nonlinearity of a Markovian point
process can help imitate the statistical properties of a simple non--Markovian
point process.

Similar inquiry was conducted by \cite{Kazakevicius2021PRE,Eliazar2021JPA}
from the perspective of self--similar stochastic processes. Self--similarity
of a stochastic process implies that the process is invariant (in
the statistical sense) to proper scaling of time and space axes \cite{Kunsch2020}.
\cite{Kazakevicius2021PRE} has explored how the nonlinear transformations
of time and space axes affect the dynamics of the noisy voter model.
While \cite{Eliazar2021JPA} has shown that there is only one Ito
diffusion that is self--similar for any given Hurst exponent $H$.
Ito stochastic differential equation analyzed in \cite{Eliazar2021JPA}
belongs to the same class of stochastic differential equation as the
ones arising from our approach \cite{Kazakevicius2021Entropy}. Unlike
\cite{Eliazar2021JPA}, our analysis here is numerical and focused
directly on the long--range memory properties of the processes. Thus
we draw direct comparisons between a nonlinear Markovian process and
a non--Markovian process.

This paper is organized as follows. A non--Markovian point process
based on the fractional Brownian motion is introduced and analyzed
from the numerical perspective in Section~\ref{sec:fractional-point-process}.
In Section~\ref{sec:markovian-point-process}, we derive a simple
Markovian point process generating the same distribution of inter--event
times. Then in Section~\ref{sec:nonlinear-point-process}, nonlinearity
is introduced into the Markovian point process. With this correction,
we are able to replicate the power--law scaling behavior of the analyzed
statistical properties of the fractional point process. Obtained numerical
results and their implications are discussed in Section~\ref{sec:conclusions}.
The code used to obtain the reported results is available via the
GitHub repository\footnote{\url{https://github.com/akononovicius/fractional-point-process}}.

\section{Non--Markovian fractional point process\label{sec:fractional-point-process}}

The intensity of a process described by event times $t_{k}$ can be
obtained by summing over their profiles:
\begin{equation}
I\left(t\right)=\sum_{k}A_{k}\left(t-t_{k}\right).
\end{equation}
We can often assume that events happen almost instantaneously, $A_{k}\left(x\right)=a_{k}\delta\left(x\right)$,
as the profiles are usually narrow in comparison to the length of
the time series. Often modeled systems are driven by the flow of identical
or very similar elementary objects, such as electrons, TCP/IP packets,
or trades. Therefore we can assume that the modeled events are unit
events, $a_{k}=1$.

When dealing with unit point events, it is convenient to introduce
event count series as an alternative to intensity series:

\begin{equation}
N\left(t\right)=\sum_{k}\mathbf{1}_{t\leq t_{k}<t+\Delta t}.
\end{equation}
In the above $\mathbf{1}_{c}$ is the indicator function, which attains
a value of $1$ if the condition $c$ is satisfied and is zero otherwise.
While $\Delta t$ is the time window within which events are counted.
Here, let us consider a unit time window, $\Delta t=1$.

Typically describing a temporal pattern by a set of event times $\left\{ t_{k}\right\} $
is equivalent to the description by a set of inter--event times $\left\{ \tau_{k}=t_{k+1}-t_{k}\right\} $
as the dynamics of many processes do not depend on the absolute time.
Inter--event times are far more convenient from the statistical analysis
and modeling point of view.

Let us consider a point process whose inter--event time $\tau_{k}$
evolves according to the iterative equation:
\begin{equation}
\tau_{k+1}=\tau_{k}+\sigma\varepsilon_{k}^{\left(H\right)}.\label{eq:frac-pp}
\end{equation}
In the iterative equation above $\sigma$ controls the rate of change
of the inter--event times, while $\varepsilon_{k}^{\left(H\right)}$
are the samples of the fractional Gaussian noise with Hurst index
$H$, zero mean, and unit variance. We obtain samples of the fractional
Gaussian noise by using the approximate circulant method \cite{Dieker2003fbm}.

It is important to note that Eq.~(\ref{eq:frac-pp}) is a generalization
of a nonlinear Markovian point process \cite{Kaulakys1998PRE,Kaulakys1999PLA,Kaulakys2005PhysRevE}.
To simplify the analysis of this fractional point process we have
removed the drift term (effectively set $\gamma=0$ in the original
model) and nonlinearity (set $\mu=0$) with the hope that the noise
term with $H\neq\frac{1}{2}$ would cause drift and nonlinearity effects
on its own.

By definition, inter--event time can't be negative. Therefore diffusion
of inter--event times has to be restricted to the positive values.
Furthermore, the process also needs to be stationary. Thus another
boundary condition has to be introduced from the side of the larger
values. To do this, we use two inelastic reflective boundary conditions,
which limit possible values of $\tau$ to $\left[\tau_{\mathrm{min}},\tau_{\mathrm{max}}\right]$:
\begin{equation}
\tau_{k+1}=\begin{cases}
\tau_{\mathrm{min}} & \text{if }\tau_{k}+\sigma\varepsilon_{k}^{\left(H\right)}<\tau_{\mathrm{min}},\\
\tau_{\mathrm{max}} & \text{if }\tau_{k}+\sigma\varepsilon_{k}^{\left(H\right)}>\tau_{\mathrm{max}},\\
\tau_{k}+\sigma\varepsilon_{k}^{\left(H\right)} & \text{otherwise}.
\end{cases}\label{eq:frac-pp-bound}
\end{equation}
Without loss of generality, let us set $\tau_{\mathrm{min}}=b$ and
$\tau_{\mathrm{max}}=1-b$ (here $b$ is some small positive real
number). In our numerical simulations we have used $b=\sigma$.

The fractional point process considered here shouldn't be confused
with so--called fractional Poisson processes \cite{Repin2000RQE,Laskin2003CNSNS,Beghin2009EJProb,Gorenflo2015Ax,Michelitsch2020PhysA}.
In the fractional Poisson process, $\tau_{k}$ are independent and
identically distributed. The apparent non--Markovian behavior emerges
not due to inherent correlations (like those accounted for in fractional
Brownian motion) but due to Mittag--Leffler heavy--tailed inter--event
time distribution. As Mittag--Leffler distribution is a solution
of the fractional Kolmogorov--Feller equation, this Poisson process
is referred to as the ``fractional'' Poisson process. For similar
reasons apparent non--Markovian behavior emerges in the nonlinear
point process we will consider in a later section.

To find the stationary distribution of Eq.~(\ref{eq:frac-pp-bound}),
let us note that the iterative equation describes the confined fractional
Brownian motion. The stationary distribution of this process was already
considered from the numerical perspective in \cite{Guggenberger2019NJP,Vojta2020PRE}.
In \cite{Guggenberger2019NJP,Vojta2020PRE} it was shown that: for
$H=0.5$ stationary distribution is uniform (the only case which can
be derived analytically), while for $H<0.5$ and $H>0.5$, probability
depletion and accretion zones (respectively) form close to the boundaries.
In fact, \cite{Guggenberger2019NJP,Vojta2020PRE} reports power--law
scaling behavior of the stationary distribution:
\begin{equation}
p\left(w_{k}\right)\propto w_{k}^{\alpha},\quad\alpha=\frac{1}{H}-2.\label{eq:frac-pdf-scaling}
\end{equation}
In the above, $w_{k}$ is the distance to the nearest boundary condition
(i.e., $w_{k}=\min\left(\tau_{k}-\tau_{\mathrm{min}},\tau_{\mathrm{max}}-\tau_{k}\right)$).
This power--law scaling behavior is observed for $w_{k}>\sigma$,
while for $w_{k}<\sigma$, the exact form of the stationary distribution
depends on the type of the boundary condition used \cite{Guggenberger2019NJP,Vojta2020PRE}.

\begin{figure}
\begin{centering}
\includegraphics[width=0.9\textwidth]{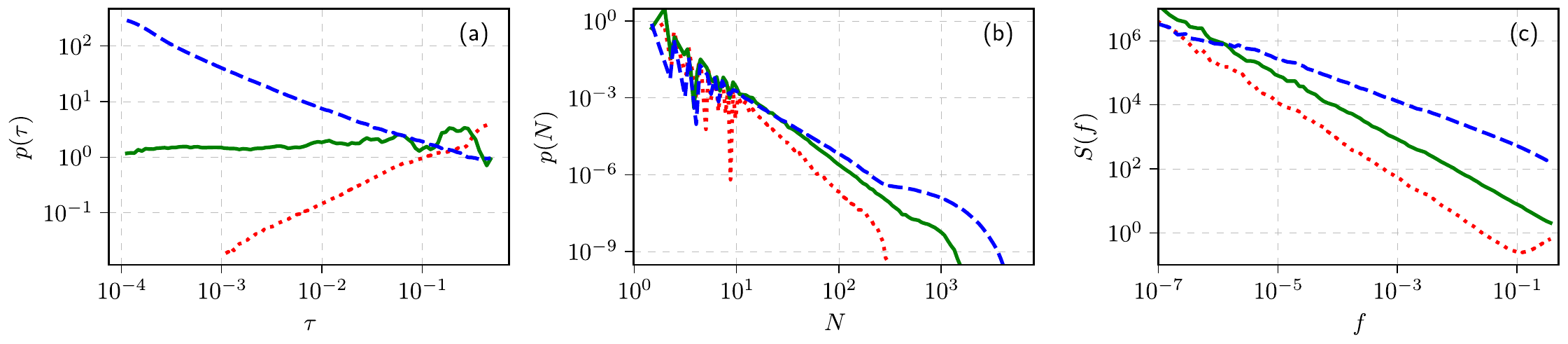}
\par\end{centering}
\caption{Statistical properties of the fractional point process: (a) inter--event
time PDF, (b) event count per unit time window PDF and (c) PSD of
the event count time series. Model parameters: $H=0.35$ and $\sigma=10^{-3}$
(red dotted line), $H=0.5$ and $\sigma=10^{-4}$ (green solid line),
$H=0.8$ and $\sigma=10^{-4}$ (blue dashed line).\label{fig:frac-pp}}
\end{figure}

In Fig.~\ref{fig:frac-pp}~(a), we see that $p\left(\tau\right)$
exhibits power--law scaling behavior as predicted by \cite{Guggenberger2019NJP,Vojta2020PRE}.
The same scaling behavior is present in the Beta distribution, particularly
by its symmetric case $\mathcal{B}e\left(\alpha+1,\alpha+1\right)$.
Beta distribution has finite support, which makes sense with the boundary
conditions we use. Thus, further we will assume that the stationary
distribution of the fractional point process is well approximated
by the symmetric Beta distribution.

Deriving analytical forms for the statistical properties of the event
count time series is an even more complicated problem. Analytical
approximations of probability density function (abbr.~PDF) and power
spectral density (abbr.~PSD) are only known for the $H=0.5$ case
\cite{Kaulakys1998PRE,Kaulakys1999PLA,Kaulakys2005PhysRevE}, while
the $H\neq0.5$ case is up to numerical exploration. Though, as can
be seen in Fig.~\ref{fig:frac-pp}~(b) and (c), both PDF and PSD
exhibit power--law scaling behavior, which appears to have different
exponents than in the $H=0.5$ case (solid green line).
The fluctuations observed in Fig.~\ref{fig:frac-pp}~(b) for small $ N $ are
numerical artifacts arising due to $ N $ being a discrete random variable.
In the following sections when estimating power--law scaling behavior of $ p
\left( N \right)$ we ignore small $ N $ value range and consider only
intermediate range, where the power--law scaling behavior is clearly observed.

\section{Simple Markovian point process\label{sec:markovian-point-process}}

Let us build a Markovian point process as a benchmark model to understand
the impact of the memory effects of the fractional Brownian motion
of the inter--event times. Our goal is to create a Markovian point
process with a similar inter--event time distribution, but driven
by a standard uncorrelated Gaussian noise instead of the fractional
Gaussian noise. Here and later on, we refer to point processes as
Markovian if the stochastic increments of inter--event times are
independent. In some of the existing literature, only Poisson point
processes are considered to be Markovian \cite{Repin2000RQE,Laskin2003CNSNS,Beghin2009EJProb,Gorenflo2015Ax,Michelitsch2020PhysA}.

For processes driven by stochastic differential equations, stationary
distributions can be easily obtained \cite{Gardiner2009Springer}.
Stationary distribution of stochastic differential equation:
\begin{equation}
\rmd\tau_{k}=\sigma^{2}f\left(\tau_{k}\right)\rmd k+\sigma\rmd W_{k},\label{eq:sde-gen}
\end{equation}
where $W_{k}$ is the standard Wiener process, is given by:
\begin{equation}
p\left(\tau_{k}\right)=\exp\left[2\int^{\tau_{k}}f\left(u\right)\rmd u\right].\label{eq:sde-gen-pdf}
\end{equation}
Let us rearrange (\ref{eq:sde-gen-pdf}) to obtain an expression that
would allow us to derive the drift function $f\left(\tau_{k}\right)$
based on the desired stationary distribution:
\begin{equation}
f\left(\tau_{k}\right)=\frac{1}{2p\left(\tau_{k}\right)}\frac{\rmd}{\rmd\tau_{k}}p\left(\tau_{k}\right).
\end{equation}
If $p\left(\tau_{k}\right)$ corresponds to PDF of the symmetric Beta
distribution, then the drift function:
\begin{equation}
f\left(\tau_{k}\right)=\frac{\alpha}{2}\left(\frac{1}{\tau_{k}}-\frac{1}{1-\tau_{k}}\right)=\left(\frac{1}{2H}-1\right)\left(\frac{1}{\tau_{k}}-\frac{1}{1-\tau_{k}}\right).
\end{equation}
The obtained stochastic differential equation can be solved numerically
by iterating the following iterative equation:
\begin{equation}
\tau_{k+1}=\tau_{k}+\sigma^{2}\left(\frac{1}{2H}-1\right)\left(\frac{1}{\tau_{k}}-\frac{1}{1-\tau_{k}}\right)+\sigma\xi_{k}.\label{eq:markov-pp}
\end{equation}
In the above, $\xi_{k}$ are uncorrelated samples from standard Gaussian
distribution (i.e, $\xi_{k}=\varepsilon_{k}^{\left(0.5\right)}$).
The smaller $\sigma$, the better the approximation of (\ref{eq:sde-gen})
by (\ref{eq:markov-pp}) will be. Thus as $\sigma$ gets smaller stationary
distribution of (\ref{eq:markov-pp}) will become more similar to
the stationary distribution of (\ref{eq:frac-pp}).

\begin{figure}
\begin{centering}
\includegraphics[width=0.3\textwidth]{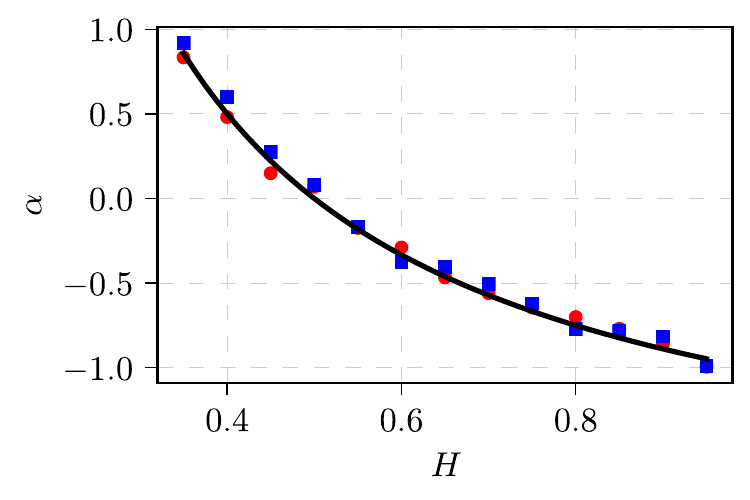}
\par\end{centering}
\caption{Power--law exponent of the PDF of inter--event times obtained by
numerically evaluating the fractional point process (red circles)
and the Markovian point process (blue squares). Black curve corresponds
to (\ref{eq:frac-pdf-scaling}). Other model parameters: $\sigma=10^{-3}$
(for the simulations of the fractional point process with $H=0.35$
and $0.4$) or $10^{-4}$ (in all other cases).\label{fig:tau_pdf}}
\end{figure}

By comparing these models, we can understand the impact of the fractional
diffusion itself, eliminating the effect of different inter--event
time distributions. Indeed (see Fig.~\ref{fig:tau_pdf}), similar
exponents $\alpha$ are obtained from both models for $H\in\left[0.35,0.95\right]$.
It means that the inter--event time distributions of both models
are similar. Furthermore, the obtained exponents $\alpha$ follow
relation (\ref{eq:frac-pdf-scaling}) reasonably well.

For the Markovian point process, unlike for the fractional model,
it is possible to derive analytical approximations of the statistical
properties of the event count time series. In this task, particularly
useful is the similarity of (\ref{eq:markov-pp}) to the iterative
equation of the point process considered in \cite{Kaulakys1998PRE,Kaulakys1999PLA,Kaulakys2005PhysRevE}:
\begin{equation}
\tau_{k+1}=\tau_{k}+\sigma^{2}\gamma\tau_{k}^{2\mu-1}+\sigma\tau_{k}^{\mu}\xi_{i}.\label{eq:old-pp}
\end{equation}
In the limit of short inter--event times ($\tau\rightarrow\tau_{\mathrm{min}}$
or $\tau\rightarrow0$) with $\mu=0$ and $\gamma=\frac{1}{2H}-1$,
the iterative equations match:
\begin{equation}
\tau_{k+1}=\tau_{k}+\sigma^{2}\left(\frac{1}{2H}-1\right)\frac{1}{\tau_{k}}+\sigma\xi_{k}.
\end{equation}
For (\ref{eq:old-pp}), it is known that event count PDF exhibits
power--law scaling behavior \cite{Gontis2004PhysA343}:
\begin{equation}
p\left(N\right)\propto N^{-\lambda},\quad\lambda=2\left(\gamma-\mu\right)+3=\frac{1}{H}+1.\label{eq:predict-lambda}
\end{equation}
Similarly, PSD of the event count time series is known to have a range
of frequencies in which it is well approximated by a power--law function
\cite{Gontis2004PhysA343}:
\begin{equation}
S\left(f\right)\propto f^{-\beta},\quad\beta=1+\frac{2\left(\gamma-\mu\right)}{3-2\mu}=\frac{1}{3}\left(\frac{1}{H}+1\right).\label{eq:predict-beta}
\end{equation}
These approximations should hold well for small $ \sigma $ and uncorrelated
$ \xi_k $ \cite{Gontis2004PhysA343, Kaulakys2005PhysRevE, Ruseckas2016JStat}.

\begin{figure}
\begin{centering}
\includegraphics[width=0.6\textwidth]{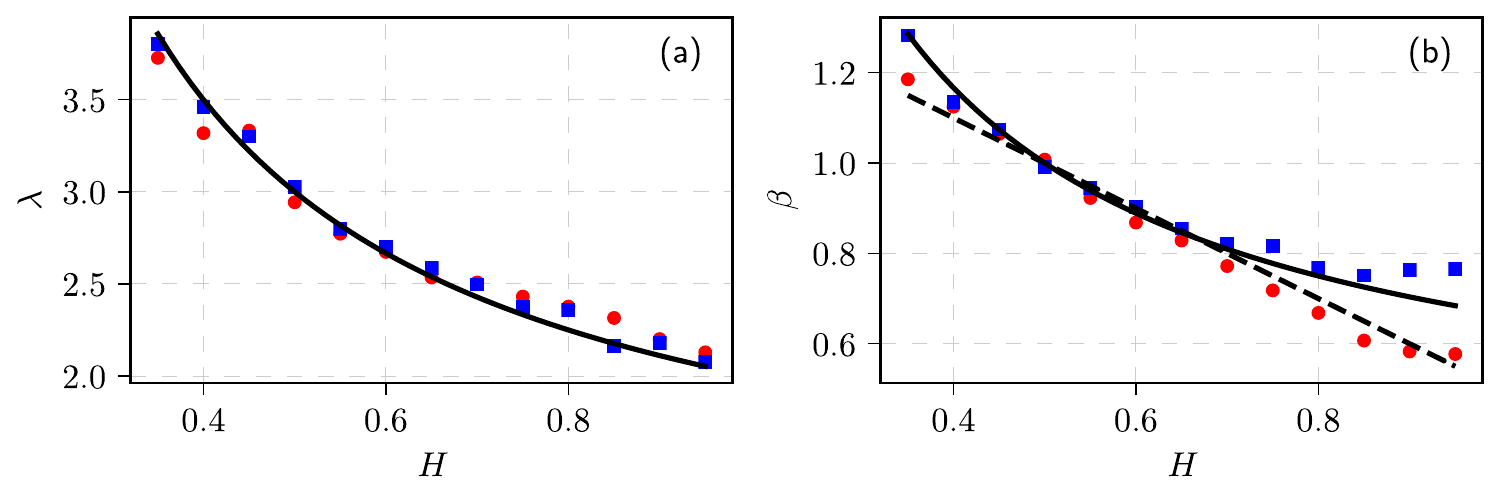}
\par\end{centering}
\caption{Power--law exponents of event count PDF (a) and PSD (b) obtained
by numerically evaluating the fractional point process (red circles)
and the Markovian point process (blue squares). Black curves follow
(\ref{eq:predict-lambda}) and (\ref{eq:predict-beta}) respectively.
Black dashed curve follows (\ref{eq:predict-beta-frac}). Model parameters
are the same as in Fig.~\ref{fig:tau_pdf}.\label{fig:event_stats}}
\end{figure}

As expected, (\ref{eq:predict-lambda}) and (\ref{eq:predict-beta})
predict the power--law exponents of the Markovian point process rather
well (see Fig.~\ref{fig:event_stats}). For the fractional point
process, exponent $\lambda$ behaves as expected for the Markovian
point process, while exponent $\beta$ attains notably different values
(with the difference getting bigger further away from $H=0.5$). It
appears that exponent $\beta$ of the fractional point process changes
linearly with $H$:
\begin{equation}
\beta\approx\frac{3}{2}-H.\label{eq:predict-beta-frac}
\end{equation}

\section{Nonlinear Markovian point process\label{sec:nonlinear-point-process}}

Using a simple Markovian point process driven by iterative equation
(\ref{eq:markov-pp}) we were able to mimic the influence of $H$
on exponents $\alpha$ and $\lambda$ of the fractional point process
driven by iterative equation (\ref{eq:frac-pp}). Let us introduce
a correction to the Markovian point process that would allow us to
reproduce $\beta$ values observed for the fractional point process.
Let us follow in the footsteps of \cite{Kaulakys1998PRE,Kaulakys1999PLA,Kaulakys2005PhysRevE}
and add nonlinearity into the Markovian point process.

Earlier, we have assumed that the diffusion function of the Markovian
point process is equal to unity. If we want to introduce nonlinearity,
we will also have to include it in the diffusion function. If the
diffusion function is not unity, but some other function $g\left(\tau_{k}\right)$,
then the corresponding drift function \cite{Gardiner2009Springer}:
\begin{equation}
f\left(\tau_{k}\right)=\frac{g^{2}\left(\tau_{k}\right)}{2p\left(\tau_{k}\right)}\frac{\rmd}{\rmd\tau_{k}}p\left(\tau_{k}\right)+g\left(\tau_{k}\right)\frac{\rmd}{\rmd\tau_{k}}g\left(\tau_{k}\right),\label{eq:nonlin-drift}
\end{equation}
here $p\left(\tau_{k}\right)$ would describe the PDF of the desired
stationary distribution. In our particular case, $p\left(\tau_{k}\right)$
is the PDF for the symmetric Beta distribution.

For $g\left(\tau_{k}\right)$, infinitely many different choices would
be valid, though they are suitable for our purposes only as long as
$g\left(\tau_{k}\right)\propto\tau_{k}^{\mu}$ in the limit of short
inter--event times. Consider the simplest possible alternative:
\begin{equation}
g\left(\tau_{k}\right)=\tau_{k}^{\mu}.\label{eq:nonlin-diff}
\end{equation}
Power--law functions were also considered in \cite{Eliazar2021JPA} with a
goal to produce Markovian process with self--similarity property.
Putting Eq.~(\ref{eq:nonlin-diff}) and the PDF for the symmetric
Beta distribution into Eq.~(\ref{eq:nonlin-drift}) yields:
\begin{equation}
f\left(\tau_{k}\right)=\frac{1}{2}\left(\frac{\alpha+2\mu}{\tau_{k}}-\frac{\alpha}{1-\tau_{k}}\right)\tau_{k}^{2\mu}.
\end{equation}

The nonlinear Markovian point process,

\begin{equation}
\tau_{i+1}=\tau_{i}+\frac{\sigma^{2}}{2}\left(\frac{\frac{1}{H}-2+2\mu}{\tau_{i}}-\frac{\frac{1}{H}-2}{1-\tau_{i}}\right)\tau_{i}^{2\mu}+\sigma\text{\ensuremath{\tau_{i}^{\mu}}}\xi_{i},\label{eq:nonlin-markov-pp}
\end{equation}
should reproduce all three consider exponents of the fractional point
process. Note that this process also retains the similarity to the
nonlinear point process considered in \cite{Kaulakys1998PRE,Kaulakys1999PLA,Kaulakys2005PhysRevE}.
In the limit of short inter--event times ($\tau\rightarrow\tau_{\mathrm{min}}$)
and with $\gamma=\frac{1}{2H}-1+\mu$ this iterative equation reduces
to (\ref{eq:old-pp}). Therefore we can reasonably expect that for
(\ref{eq:nonlin-markov-pp}), the relationships between model parameters
and the power--law exponents would still hold \cite{Kaulakys1998PRE,Kaulakys1999PLA,Kaulakys2005PhysRevE}:
\begin{equation}
\alpha=2\left(\gamma-\mu\right),\quad\lambda=\alpha+3,\quad\beta=1+\frac{\alpha}{3-2\mu}.
\end{equation}

To reproduce $\beta_{F}\left(H\right)$ observed for the fractional
point process (hence the index $F$) we need to figure out how power--law
exponent $\mu\left(H\right)$ of the nonlinear Markovian point
process depends on the Hurst parameter. This is done by solving:
\begin{equation}
\beta_{F}\left(H\right)=1+\frac{\alpha\left(H\right)}{3-2\mu\left(H\right)}.\label{eq:beta-hurst}
\end{equation}
For $\beta_{F}\left(H\right)\neq1$ the solution of Eq.~(\ref{eq:beta-hurst})
is given by:
\begin{equation}
\mu\left(H\right)=\frac{3}{2}+\frac{\alpha\left(H\right)}{2\left(1-\beta_{F}\left(H\right)\right)}=\frac{3}{2}+\frac{\frac{1}{H}-2}{2\left(1-\beta_{F}\left(H\right)\right)}.\label{eq:mu-general-form}
\end{equation}
Unfortunately, with Eq.~(\ref{eq:predict-beta-frac}) as $\beta_{F}\left(H\right)$
the value of $\mu$ in the vicinity of $H=\frac{1}{2}$ is close to
$-\frac{1}{2}$, while from (\ref{eq:frac-pp}) it is evident that
it should be $0$. To obtain the proper value of $\mu$ in the vicinity
of $H=\frac{1}{2}$ we would need to change the slope of dependence
on $H$:
\begin{equation}
\beta_{F}\left(H\right)=1-\frac{4}{3}\left(H-\frac{1}{2}\right).
\end{equation}
While this now yields the proper value of $\mu$ close to $H=\frac{1}{2}$,
predictions of $\beta_{F}$ for other $H$ noticeably disagree with
the results of the numerical simulation. Though the agreement can
be improved by including second--order term:
\begin{equation}
\beta_{F}\left(H\right)=1-\frac{4}{3}\left(H-\frac{1}{2}\right)+C_{2}\left(H-\frac{1}{2}\right)^{2}.\label{eq:predict-beta-frac-2}
\end{equation}
The best fit to the numerical simulation results is obtained with
$C_{2}\approx\frac{2}{3}$. Then Eq.~(\ref{eq:mu-general-form})
becomes:
\begin{equation}
\mu\left(H\right)=\frac{3}{2}-\frac{3}{5H}-\frac{6}{5\left(5-2H\right)}.\label{eq:mu-form}
\end{equation}

\begin{figure}
\centering{}\includegraphics[width=0.6\textwidth]{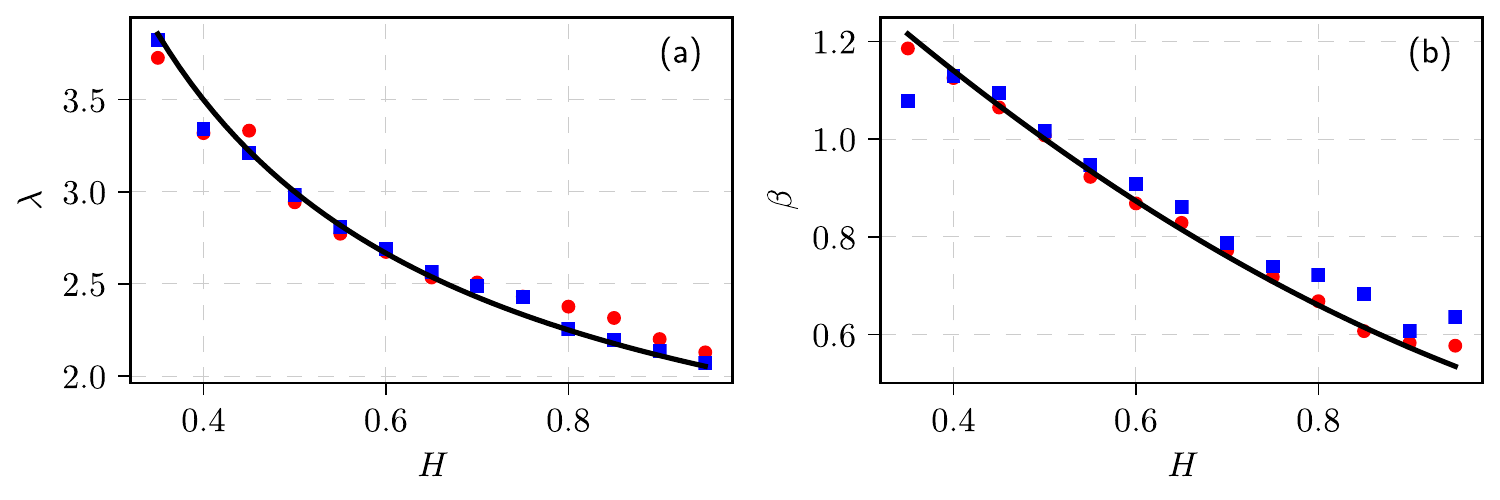}\caption{Power--law exponents of event count PDF (a) and PSD (b) obtained
by numerically evaluating the fractional point process (red circles)
and the nonlinear Markovian point process (blue squares). Black curves
follow (\ref{eq:predict-lambda}) and (\ref{eq:predict-beta-frac-2}).
Fractional point process data is the same as in Fig.~\ref{fig:event_stats}.
Nonlinear Markovian point process was simulated with $\sigma=10^{-3}$
(for simulations with $H=0.35$) and $10^{-4}$ (in all other cases).\label{fig:event_stats_adv}}
\end{figure}

Numerical simulation of the nonlinear Markovian point process, Eq.~(\ref{eq:nonlin-markov-pp}),
with $\mu$ determined by Eq.~(\ref{eq:mu-form}), produces power--law
exponents similar to the ones obtained by numerically simulating the
fractional point process. As seen in Fig.~\ref{fig:event_stats_adv}
introduction of nonlinearity has enabled reasonable reproduction of
the PSD exponent $\beta$. Though the numerical simulation results
slightly deviate from the analytical prediction. The agreement between
analytical and numerical results of the nonlinear Markovian point
process should improve with smaller $\sigma$
\cite{Gontis2004PhysA343,Kaulakys2005PhysRevE,Ruseckas2016JStat},
yet decreasing $\sigma$
increases the required length of generated time series, the real simulation
time, and required computer memory resources.

\section{Conclusions\label{sec:conclusions}}

Here we have introduced a generalization of the point process proposed
in \cite{Kaulakys1998PRE,Kaulakys1999PLA,Kaulakys2005PhysRevE} by
replacing the Brownian motion of the inter--event times with fractional
Brownian motion. Thus, by construction, our generalization exhibits
long--range memory stemming from the persistence or the anti--persistence
of inter--event times. We have numerically explored this model from
the perspective of its long--range memory properties.

The considered fractional point process can generate time series exhibiting
power--law scaling in PDF and PSD. Time series generated with persistent
inter--event times (using fractional Brownian motion with $H>0.5$)
exhibit anti--persistent behavior (PSD exponent $\beta<1$), while
time series with anti--persistent inter--event times ($H<0.5$)
appear to exhibit persistent behavior ($\beta>1$). PDF exponent $\lambda$
also gets smaller as Hurst exponent $H$ of the underlying fractional
Brownian motion grows larger. Although, we weren't able to numerically
observe the behavior of the fractional point process for the whole
range of $H$. For most cases driven by the anti--persistent fractional
Brownian motion the fractional point process doesn't explore the phase
space quick enough, the lower $H$ the longer time series are needed
to properly observe the long--term behavior of the process. Cases
with $H<0.35$ were too demanding for the computational resources
and took too long to run.

The observed power--law scaling behavior of the generated time series
PDF can be replicated by a simple Markovian point process mimicking
the inter--event time PDF of the fractional point process. Based
on the prior numerical exploration of the confined fractional Brownian
motion \cite{Guggenberger2019NJP,Vojta2020PRE}, we have assumed that
the inter--event time distribution of the considered fractional point
process is the symmetric Beta distribution. From the numerical simulations,
we can conclude that this assumption is reasonable as generated inter--event
time PDFs appear similar in both models. Event count (time series)
PDFs appear similar as well.

Although, the reproduction of the observed power--law scaling behavior
in the PSD requires the introduction of nonlinearity into the Markovian
point process. We have considered the simplest case, when the diffusion
term of the point process is a power--law function of the prior inter--event
time. Using the nonlinear Markovian point process with symmetric Beta
inter--event time distribution, we have reproduced the power--law
scaling behavior in PDF and PSD observed in the fractional point process.

We have introduced two similar point processes -- one driven by a
fractional noise and the other by a white noise. The latter can mimic
the former due to nonlinearity in drift and diffusion terms of process.
Having such two similar and comparable models of differing nature
should allow building, testing and, tuning indicators attempting to
differentiate between the ``true'' and spurious long--range memory
processes. Some of the more recent attempts to discriminate between
two different types of memory include \cite{Eliazar2013PhysRep,Spiechowicz2017SciRep,Masoliver2017EPJB,Wang2018ModPhysLettB,Liang2019AMR,Yang2020CompDiff,Gontis2022CNSNS,Eliazar2022JPA}.

{\small
\section*{Author contributions}

Aleksejus Kononovicius: Conceptualization, Methodology, Software, Writing --
Original Draft, Writing -- Review \& Editing, Visualization. Rytis
Kazakevi\v{c}ius: Conceptualization, Methodology, Writing -- Review \& Editing.
Bronislovas Kaulakys: Conceptualization, Writing -- Review \& Editing,
Supervision.
}


\end{document}